\documentstyle[12pt,epsfig,amsmath,amssymb,pstricks]{article}
\begin{document}

\title{\bf A model-theoretic interpretation \\ of \\ environmentally-induced superselection}

\author{{Chris Fields}\\ \\
{\it 815 East Palace \# 14}\\
{\it Santa Fe, NM 87501 USA}\\ \\
{fieldsres@gmail.com}}
\maketitle

\begin{abstract}
The question of what constitutes a ``system'' is foundational to quantum measurement theory.  Environmentally-induced superselection or ``einselection'' has been proposed as an observer-independent mechanism by which apparently classical systems ``emerge'' from physical interactions between degrees of freedom described completely quantum-mechanically.  It is shown here that einselection can only generate classical systems if the ``environment'' is assumed \textit{a priori} to be classical; einselection therefore does not provide an observer-independent mechanism by which classicality can emerge from quantum dynamics.  Einselection is then reformulated in terms of positive operator-valued measures (POVMs) acting on a global quantum state.  It is shown that this re-formulation enables a natural interpretation of apparently-classical systems as virtual machines that requires no assumptions beyond those of classical computer science.
\end{abstract}


\section{Introduction}

The concept of environmentally-induced superselection or ``einselection'' was introduced into quantum mechanics by W. Zurek \cite{zurek:81, zurek:82} as a solution to the ``preferred-basis problem,'' the problem of selecting one from among the arbitrarily many possible sets of basis vectors spanning the Hilbert space of a system of interest.  Zurek's solution was brilliantly simple: it transferred the preferred basis problem from the observer to the environment in which the system of interest is embedded, and then solved the problem relative to the environment.  Any physical system $S$ interacts with the environment $E$ in which it is embedded; this interaction can be represented by a Hamiltonian $H_{S-E}$.  If the internal states of both $S$ and $E$ are stable over the instantaneous timeframes relevant to individual observations, then the instantaneous time evolution of the joint system-environment state $|S \otimes E\rangle$ is detemined only by $H_{S-E}$, i.e. the Schr\"odinger equation $(\partial / \partial t)|S \otimes E\rangle = -\imath \hbar H_{S-E}|S \otimes E\rangle$ holds.  If this is the case, both $|S\rangle$ and $|E\rangle$ must be eigenstates of $H_{S-E}$, so there are eigenbases $\lbrace |s_{i}\rangle \rbrace$ and $\lbrace |e_{i}\rangle \rbrace$ for $S$ and $E$ respectively with basis vectors $|s_{k}\rangle \in \lbrace |s_{i}\rangle \rbrace$ and $|e_{j}\rangle \in \lbrace |e_{i}\rangle \rbrace$ such that $\langle e_{j}| H_{S-E} |s_{k}\rangle = \lambda_{jk}$ where $\lambda_{jk}$ is a real number representing the instantaneous coupling between $S$ and $E$.  The basis $\lbrace |s_{i}\rangle \rbrace$ for $S$ that is selected by $H_{S-E}$ is completely observer-independent; to employ a canonical example, a momentarily-stable ambient visible-spectrum photon field einselects the position basis for all momentarily-stable systems significantly larger than a few microns, completely independently of whether anyone interacts with the photon field by eye.  

Einselection revolutionized decoherence theory by converting the environment from a passive sink for quantum coherence into an active ``witness'' that continuously determines the observable states of the systems embedded in it \cite{zurek:98, zurek:03, zurek:04, schloss:04, zurek:05, schloss:07}.  Quantum states that are continually ``witnessed'' by the environment are effectively classical; hence it has been proposed, under the rubric of ``quantum Darwinism,'' that einselection provides an observer-independent physical mechanism for the emergence of classicality \cite{zurek:06, zurek:09}.  If this proposal is correct, einselection answers one of the deepest questions in quantum measurement theory: the question 
of ``how one can define systems given an overall Hilbert space `of everything' and the total Hamiltonian'' (\cite{zurek:98} p. 1794).  Hence it also answers Einstein's famous question of whether ``the Moon exists only when I look at it'' (\cite{pais:79}; quoted in \cite{mermin:85} p. 38); nobody has to look, because the \textit{environment} is always looking (cf. \cite{wheeler:75, zurek:02}).   In recognition of this, einselection has been called ``the most important and powerful idea in quantum theory since entanglement'' (\cite{landsman:07} p. 94).

The present paper shows that a recognizably classical world of discrete, bounded objects acted upon by external forces only emerges via einselection if ``systems'' are defined in an observer-dependent way.  It thus shows that einselection does not provide an observer-\textit{independent} mechanism for the emergence of classicality.  It then describes an alternative view of einselection in which positive operator-valued measures (POVMs; e.g. \cite{nielsen-chaung:00}, Ch. 2), not system-environment interaction Hamiltonians, provide the information with which unique bases for discrete, observer-recognizable systems are selected.  This alternative view is explicitly observer-dependent, and relies on the idea that the environment $E$ implements an information channel that observers employ to investigate the properties of systems of interest.  In this alternative view, einselection is not as a physical process, but a semantic process that selects a particular classical virtual machine from among the arbitrarily many that are consistent with the quantum information dynamically encoded by the environment.

The paper is organized as follows.  The next section, ``Observer-independent versus observer-dependent notions of the `emergence' of classicality'' distinguishes the observer-independent notion of emergence supported by the environment as witness formulation of decoherence \cite{zurek:98, zurek:03, zurek:04, schloss:04, zurek:05, schloss:07} from the observer-dependent notion of emergence advocated by Wallace \cite{wallace:05} among others.  The third section, ``Observer-independent einselection does not produce classicality'' shows that einselection does not provide a mechanism for the observer-independent emergence of classicality by showing that the ``environment as witness'' formalism implicitly assumes the effective classicality of the environment.  The fourth section, ``Reformulating einselection using POVMs'' develops the necessary formalism for viewing einselection as a consequence of interactions between observers and physically-implemented information channels.  The fifth section, ``Systems as virtual machines'' shows that einselection as reformulated in terms of POVMs can be viewed as an entirely semantic process.  From this perspective, einselection is formally analogous to the process of interpreting the observed behavior of any physical device as classical computation, i.e. as the realization of a classical virtual machine.  The paper concludes by suggesting that this semantic view of einselection may provide a precise and self-consistent way of describing the initialization of quantum computing algorithms with classical data.

\section{Observer-independent versus observer-dependent \\ notions of the ``emergence'' of classicality}

Before proceeding, it is necessary to say with some precision what is meant by ``classicality'' and to distinguish ``observer-independent'' from ``observer-dependent'' notions of its ``emergence.''   Let us explicitly assume that minimal quantum mechanics - quantum mechanics with no ``collapse postulate'' - is true universally, and hence that the Schr\"odinger equation $(\partial / \partial t)|U\rangle = -\imath \hbar H_{U}|U\rangle$ describes the evolution of the universe $U$ as a whole; all available experimental evidence is consistent with this asumption (e.g. \cite{schloss:06}).  From an interpretative perspective, this assumption corresponds to ``Stance 1'' as defined by Landsman, the view that ``quantum theory is fundamental and universally valid, and the classical world has only 'relative' or 'perspectival' existence'' (\cite{landsman:07}, p. 6).  It is with respect to this assumption of the correctness of minimal quantum mechanics that einselection is defined; the goal of einselection is to render the ``perspectival'' existence of quantum systems, and in particular the macroscopic quantum systems that \textit{appear} classical to observers, relative to the ``perspective'' of the universe as a whole  \cite{zurek:98, zurek:03, zurek:04, zurek:05}.  

Let us also assume that the universe comprises an arbitrarily large and hence effectively infinite collection of ``elementary'' entities characterized by quantum degrees of freedom.  The behavior of a system $S$ that comprises some finite subset of these elementary entities is then \textit{effectively classical} if it can be described to an approximation sufficient for all practical purposes (FAPP) by the formalism of classical physics, which will for simplicity be taken to mean non-relativistic classical physics.  Classical physics requires spatially separated systems to have separable states - i.e. that there be no entanglement - and requires all measurements to commute and be non-contextual.  Schlosshauer, for example, characterizes classicality by the absence of ``quantum'' restrictions on measurement:

\begin{quote}
``Here (i.e. in classical physics) we can enlarge our `catalog' of physical properties of the system (and therefore specify its state more completely) by performing an arbitrary number of measurements of identical physical quantities, in any given order.  Moreover, many independent observers may carry out such measurements (and agree on the results) without running into any risk of disturbing the state of the system, even though they may have been initially completely ignorant of this state.''
\begin{flushright}
(\cite{schloss:07} p. 16)
\end{flushright} 
\end{quote}

The ability of ``initially completely ignorant'' observers to discover the states of a classical system is similarly emphasized by Ollivier, Poulan and Zurek in their operational definition of objectivity:

\begin{quote}
``A property of a physical system is \textit{objective} when it is:
\begin{list}{\leftmargin=2em}
\item
1. simultaneously accessible to many observers,
\item
2. who are able to find out what it is without prior knowledge about the system of interest, and 
\item
3. who can arrive at a consensus about it without prior agreement.''
\end{list}
\begin{flushright}
(p. 1 of \cite{zurek:04}; p. 3 of \cite{zurek:05})
\end{flushright}
\end{quote}

Observers can approach a system from a position of complete ignorance only if they can take its existence as a well-defined \textit{entity} for granted: picking the system out from the other furniture of the universe must be entirely unproblematical.  Being a well-defined entity requires having a well-defined boundary that enforces causal independence from other such entities.  Following Einstein, Fuchs expresses this requirement rhetorically: ``what is it that A and B are spatially distant things than that they are causally independent?'' (\cite{fuchs:10} p. 15).  From an ontological perspective, a classical Moon must simply be there, as a bounded entity distinct from everything else, whether anyone bothers to look at it or not.

Classicality in this sense can ``emerge'' from the unitary dynamics of the elementary entities of a quantum universe by an observer-independent physical mechanism only if this mechanism enforces the effective boundedness and effective separability of distinct and spatially-separated collections of elementary entities.  Ollivier, Poulan and Zurek's claim that the ``witnessing'' of the states of quantum systems by the environment renders them ``objective'' for initially-ignorant observers \cite{zurek:04, zurek:05} is the claim that einselection is such a mechanism.  For this emergence claim to be true, einselection must not only establish a preferred basis for $S$, but also establish a boundary for $S$ that separates and hence distinguishes it from all spatially-distinct systems.  This boundary need not be ontological in the strict sense of being an entity in addition to the elementary entities composing the universe, but must be ``FAPP ontological'' in the sense that an ``initially completely ignorant'' observer can take it for granted.

This FAPP ontological requirement for the observer-independent emergence of classicality can be understood by contrast with the much less stringent requirements for observer-\textit{dependent} emergence, the emergence of an object, property or phenomenon by means of a mechanism that depends on particular properties of a specified observer or class of observers.  A particularly clear example of observer-dependent emergence is provided by Wallace, who defines an emergent ``pattern'' as follows:

\begin{quote}
``A macro-object is a pattern, and the existence of a pattern as a real thing depends on the usefulness -
in particular, the explanatory power and predictive reliability - of theories which admit that pattern in their ontology.''
\begin{flushright}
(p. 93 of \cite{wallace:05}; p. 58 of \cite{wallace:10})
\end{flushright}
\end{quote}

This definition relies on the ``usefulness'' of theories, and hence on the existence of observers who find them useful.  While Wallace describes decoherence and einselection using the standard observer-independent formalism, he notes that the identification of einselected systems as quasi-classical requires ``making a fairly unprincipled choice of system-environment split and then noticing that that split led to interesting behaviour'' (\cite{wallace:10} p. 64).  Here the mechanism of einselection is not required to establish the boundary separating the system from its environment; this boundary is \textit{stipulated} by an observer, who then ``notices'' that the behavior of the stipulated system is ``interesting.''  Thus while Wallace never employs the term ``observer-dependent'' and never discusses the properties that an observer might need to possess in order to find behavior interesting or engage in the construction of useful theories, his account of emergence depends on the existence of observers having these properties.  

If the system-environment boundary at which $H_{S-E}$ is defined can only be specified in an observer-dependent way, then the emergence of classicality by einselection cannot be regarded as observer-independent: the Moon is only there if someone \textit{stipulates} its boundary, and any such stipulation requires looking.  ``Emergence'' in this observer-dependent sense is not ontological even FAPP; to say that something is ``emergent'' in this sense is rather to make an epistemic claim that its behavior is unpredictable FAPP from fundamental physics.  For Wallace, such epistemic claims of emergence apply equally to phonons, tigers, and chess-playing programs.  On Wallace's observer-dependent account of emergence, if no observer found the Moon to be interesting, there would still be a dense concentration of elementary particles in the sky, but there would be no emergent quasi-classical object.

Wallace's implicit characterization of observers as theory-constructors with an ability to recognize some phenomena as interesting can be contrasted with the more typical characterization of an observer as ``any physical system having a definite state of motion'' (\cite{rovelli:96} p.  1641) or ``a quantum system interacting with the observed system'' (\cite{schloss:07} p. 361).  These traditional ``Galilean'' characterizations do not attribute any particular \textit{structure} to observers, and are hence consistent with observers being ``initially completely ignorant'' when they encounter a system to be observed.  In particular, a Galilean observer does not possess ``pragmatic information'' \cite{roederer:05} about what properties may or may not be interesting, and does not actively \textit{identify} particular systems as targets of interaction.  Hence the distinction between observer-dependent and observer-independent accounts of emergence can be formulated as a distinction between non-Galilean and Galilean concepts of the observer. Non-Galilean observers stipulate what systems exist and hence do not require FAPP ontological emergence; Galilean observers do not make such stipulations, have no choice but to take the existence of systems for granted, and hence do require FAPP ontological emergence.  It has been argued elsewhere that Galilean observers are not capable of identifying systems even if their existence is taken for granted \cite{fields:11b}.  The next section shows that einselection and environmental ``witnessing'' do not justify taking ``emergent'' systems for granted.

\section{Observer-independent einselection does not produce classicality}

In orthodox, realist quantum mechanics, a ``system'' is represented by a Hilbert space, each vector of which corresponds to an admissible physical state (e.g. \cite{vonNeumann:32}); the universe $U$ can, therefore, be considered to be characterized by a Hilbert space $\mathcal{H}_{\mathit{U}}$.  Any finite system $S$ can then be thought of as a proper subspace $\mathcal{H}_{\mathit{S}} \subset \mathcal{H}_{\mathit{U}}$.  In an observer-free universe, the environment $E$ of any system $S$ can be taken to be $U \setminus S$; this expression can be employed as an approximation in a universe containing only Galilean observers, since such observers are quantum systems that can be assumed to be negligibly small compared to $U$, and by assumption they do not stipulate systems, including $E$.  In this case, the $S-E$ boundary is the boundary of the subspace $\mathcal{H}_{\mathit{S}}$ within $\mathcal{H}_{\mathit{U}}$.  A Hamiltonian $H_{S-E}$ can be defined at this boundary; $H_{S-E} = 0$ if but only if the degrees of freedom within $S$ do not interact in any way with the degrees of freedom outside $S$, a situation that can be dismissed as physically uninteresting.  Provided that over a time interval $\Delta t$ the self-Hamiltonians $H_{S}$ and $H_{E}$ are small compared to $H_{S-E}$, eigenbases $\lbrace |s_{i}\rangle \rbrace$ and $\lbrace |e_{i}\rangle \rbrace$ for $S$ and $E$ respectively are einselected by $H_{S-E}$ within the interval $\Delta t$.  As the state of the environment is only observationally relevant close to both $S$ and some observer $O$, the requirement that $H_{E}$ is small can be relaxed to the requirement that $H_{F}$ is small for some proximate fragment $F \subset E$; in this case a basis $\lbrace |f_{i}\rangle \rbrace$ of $F$ is einselected.

If no physical principle restricts the choice of $S$, this construction can be carried out separately for every subspace $\mathcal{H}_{\mathit{S}}$ of $\mathcal{H}_{\mathit{U}}$ and proxmiate fragment $F \subset U \setminus S$ for which a $\Delta t$ can be found during which $H_{S}$ and $H_{F}$ are small.  Thus if no physical principle restricts the choice of $S$, the state of every subset of $U$ that is surrounded by a reasonably stable local environment can be considered to be einselected by $U$: $U$ thus ``witnesses'' the states of not one but \textit{all} of its subsystems.  This formal possibility immediately raises three questions.  First, are there subspaces of $\mathcal{H}_{\mathit{U}}$ for which the requirement of small $H_{S}$ and $H_{F}$ are satisfied over all but a negligible subset of time intervals?  Second, are there significant time intervals during which the requirements that $H_{S}$ and $H_{F}$ be small are simultaneously satisfied for multiple distinct subspaces of $\mathcal{H}_{\mathit{U}}$?  Finally, are there significant time intervals for which the requirements that $H_{S}$ and $H_{F}$ be small are simultaneously satisfied both for multiple distinct subspaces of $\mathcal{H}_{\mathit{U}}$ and for multiple environmental fragments that only intersect far from any observers?  The first condition must be satisfied, clearly, by any system $S$ that emerges as a classical object for a significant period of time.  The second condition must be satisfied by any significant time interval during which multiple classical objects emerge.  The third condition must be satisfied, within the quantum Darwinist formalism \cite{zurek:06, zurek:09}, by any significant time interval during which multiple classical objects emerge for multiple observers, as this formalism requires that the observers occupy mutually-separable environmental fragments.

Let us examine these questions in detail, beginning with the first.  Suppose $\mathcal{H}_{\mathit{S}} \subset \mathcal{H}_{\mathit{U}}$ and that $H_{S}$ and $H_{F}$ are small compared to $H_{S-E}$ and hence to $H_{S-F}$.  Since both $S$ and $F$ are by assumption composed of elementary entities, the Hamiltonian $H_{S-F} = \sum_{i,j} H_{ij}$ where $H_{ij}$ is the interaction between the $i^{th}$ elementary entity $S_{i}$ within $S$ and the $j^{th}$ elementary entity $F_{j}$ within $F$.  If the average interaction of these elementary entities is $<H_{ij}>$ and $S$ is taken to comprise $N$ elementary entities, then $N<H_{ij}> ~ >> H_{S}$ or $H_{F}$.  Consider now an alternative system $S^{\prime} = S \otimes F_{k}$, where $F_{k}$ is one of the elementary entities of $F$.  If $H_{ik} \geq~<H_{ij}>$ for $j \neq k$, then $H_{S^{\prime}} \geq H_{S}$.  If $H_{ik} \leq~<H_{ij}>$ for $j \neq k$, then $H_{F^{\prime}} \geq H_{F}$, where $F^{\prime} = F \setminus F_{k}$.  In either case, $S^{\prime}$ is less ``stable'' that $S$; hence some number of such systems $S^{\prime}$ can be constructed that satisfy the conditions for einselection, but as elementary entities are added to $S$, eventually an $S^{\prime}$ will be constructed for which either $H_{S^{\prime}}$ or $H_{F^{\prime}}$ is too large for einselection.  Similar sets of alternative systems can be constructed by subtracting elementary entities from $S$, or by adding some and subtracting other elementary entities from $S$.  In either case, some of the constructed alternatives satisfy the conditions for einselection, but too many alterations of $S$ will generate alternative systems that violate the conditions for einselection.  Satisfaction of the conditions for einselection by $S$ is, therefore, consistent with the existence of a ``halo'' of both larger and smaller systems within the environmental fragment $F$ that also satisfy the conditions for einselection, but is inconsistent with arbitrarily many alternative systems within $F$ satisfying these conditions.  Nothing in this reasoning, of course, limits the einselection of alternative systems for different choices of $F$.

As Wallace \cite{wallace:05} points out in his discussion of decoherent histories, many of the alternative systems that satisfy the conditions for einselection nearly as well as $S$, and that may be indistinguishable from $S$ by observations carried out at any finite resolution $\epsilon$, may be spatially dissociated or otherwise distinctly non-classical.  As the time interval $\Delta t$ over which $S$ and its alternatives satisfy the conditions for einselection increases, moreover, the dominance of $H_{S-E}$ over $H_{S}$ guarantees that $S$ itself will increasingly dissolve into $F$.  Hence if no physical principle restricts the choice of $S$, any system that appears classical by einselection will be accompanied by a halo of indistinguishable and in some cases overtly non-classical systems at all times, and $S$ will itself become increasingly non-classical over time.  While the existence of such a halo has no effect on the einselection of states of $S$, it does create an insurmountable problem for observers attempting to uniquely identify $S$, as has been pointed out previously \cite{fields:10, fields:11a}.

The alternative systems composing the halo of a system $S$ that are constructed as above all overlap with $S$.  It is clear, however, that non-overlapping systems embedded in the same environmental fragment $F$ cannot simultaneously satisfy the conditions for einselection.  Suppose $S^{1}$ and $S^{2}$ are embedded in $F = F^{1} \cup F^{2}$ where $F^{1} \supset S^{2}$ and surrounds $S^{1}$, and $F^{2} \supset S^{1}$ and surrounds $S^{2}$.  If $S^{1}$ and $S^{2}$ simultaneously satisfy the conditions for einselection, then $H_{S^{1}-F^{1}} >> H_{F^{1}}$ and $H_{S^{2}-F^{2}} >> H_{F^{2}}$.  This requires that $H_{S^{1}-F^{1}}$ be both much larger and much smaller than $H_{S^{2}-F^{2}}$, which is impossible.  Non-overlapping systems can, therefore, simultaneously satisfy the conditions for einselection only if they are embedded in distinct environmental fragments.  As einselection forces the state of the fragment $F^{i}$ in which $S^{i}$ is embedded to be an eigenstate of $H_{S^{i}-F^{i}}$, distinct fragments $F^{1}$ and $F^{2}$ can only overlap if their states are simultaneous eigenstates of $H_{S^{1}-F^{1}}$ and $H_{S^{2}-F^{2}}$.  If this is the case, however, $|S^{1}\rangle$ or $|S^{2}\rangle$ must also be simultaneous eigenstates of both $H_{S^{1}-F^{1}}$ and $H_{S^{2}-F^{2}}$, in which case a measurement of either $|S^{1}\rangle$ or $|S^{2}\rangle$ determines the other without signalling.  Such entanglement between $|S^{1}\rangle$ or $|S^{2}\rangle$ violates not only classicality but the \textit{appearance} of classicality; hence $S^{1}$ and $S^{2}$ can simultaneously emerge into classicality by einselection only if their respective environmental fragments $F^{1}$ and $F^{2}$ are not only distinct but separable.

The preceding discussion has implicitly assumed that only a single observer extracts information from the environment by passively interacting with $F$.  If multiple observers simultaneously extract information, they must extract it from separable environmental fragments in order to assure that their measurements do not interfere \cite{zurek:06, zurek:09}.  Multiple observers that share an objective ``classical world'' of multiple apparently-classical objects must, therefore, each have available a separable collection of mutually-separable environmental fragments $F^{i}$, one for each of the simultaneously-emergent objects.  Such mutually-separable environmental fragments are, however, themselves effectively classical objects: they have boundaries that prevent mutual entanglement and are thus FAPP ontological.  These effectively classical objects do not themselves ``emerge''; their existence must be assumed to explain the emergence of the simultaneously-einselected $S^{i}$.  Assuming the existence of effectively classical objects to explain the emergence of other effectively classical objects by einselection, however, accomplishes nothing, and without this assumption that the $F^{i}$ are effectively classical, the explanation of emergence by einselection collapses.

Einselection cannot, therefore, be viewed as an observer-independent mechanism for the emergence of classicality; doing so requires the assumption that the environment is effectively classical and is therefore circular.  This circularity cannot be rescued by a bootstrap argument, since an effectively classical environmental fragment must be assumed for every emerging object, and such an assumption must be made independently for every observer.  It also cannot be rescued by assuming a physical but observer-independent restriction on the choice of systems to be einselected.  Any such restriction - any specification of Zurek's ``axiom(o)'' that ``systems exist'' \cite{zurek:03} to an axiomatic assumption that only \textit{certain specified} systems exist - also introduces circularity, as it enforces classicality by violating the superposition principle for states of $U$.  If einselection is to be considered a physical mechanism for the emergence of classicality, therefore, it can only be considered to be an observer-\textit{dependent} mechanism.  The next section reformulates einselection as an observer-dependent process in which superselection is implemented not by the system-environment Hamiltonian but by an observer-deployed POVM.

\section{Reformulating einselection using POVMs}

Explicitly specifying a Hilbert space $\mathcal{H}_{\mathit{S}}$ is not the only way to specify a quantum system; one can also specify a system as \textit{whatever} yields a particular set of values when acted upon by a particular set of observables.  For example, \textit{whatever} has a mass of 0.511 MeV, an electric charge of -1 and spin 1/2 is an electron.  Observations can, in general, be represented as sets $\lbrace E_{i} \rbrace$ of positive semi-definite operators that sum to unity, i.e. as POVMs \cite{nielsen-chaung:00}.  A POVM is defined over a particular Hilbert space, and it is obviously circular to say that a system $S$ is \textit{whatever} yields a particular set of values when acted upon by a POVM $\lbrace E_{i} \rbrace$ defined over $\mathcal{H}_{\mathit{S}}$.  As emphasized by Zurek \cite{zurek:98, zurek:03}, however, observers typically obtain information about systems by acting on the environment.  In what follows, therefore, all POVMs are taken to be defined over $\mathcal{H}_{\mathit{U}}$, the Hilbert space of the universe as a whole.  As arbitrarily many components of a POVM can be normalized so as to yield arbitrarily small measures, arbitrarily many of the degrees of freedom of $U$ can be effectively insensitive to the action of any given POVM.  The degrees of freedom of $U$ that are given significant measures by a POVM $\lbrace E_{i} \rbrace$ can be considered a ``system'' $S^{E}$ \textit{recognized by} $\lbrace E_{i} \rbrace$.  Hence observers can be regarded as querying the \textit{universe} with a collection of POVMs, and identifying as particular \textit{systems} the collections of degrees of freedom recognized by those POVMs \cite{fields:11b}.

A von Neumann projective measurement \cite{vonNeumann:32} is a POVM $\lbrace \Pi_{i} \rbrace$ comprising an orthogonal set of components $\Pi_{i} = |e_{i}\rangle \langle e_{i}|$ where $\lbrace |e_{i}\rangle \rbrace$ is an orthonormal basis for the Hilbert space over which the measurement is defined.  Any POVM can be represented as an appropriately re-normalized subset of an ``informationally complete'' POVM that comprises a von Neumann projective measurement together with a set of projections on mixtures of the von Neumann basis vectors \cite{fuchs:02}.  Hence any component $E_{i}$ of a POVM $\lbrace E_{i} \rbrace$ is a projection onto \text{some} basis, even though the components do not, in general, all project onto the \textit{same} basis.  Because the components of a POVM are in general not orthogonal, a measurement with a POVM may result in non-negligible Born-rule probabilities $\langle U| E_{i} |U\rangle$ being associated with projections onto distinct bases.  Describing a system $S^{E}$ with a single basis during any given time interval $\Delta t$ therefore requires that only a single component of the relevant POVM $\lbrace E_{i} \rbrace$ has a significant Born-rule probability during $\Delta t$.

If within a time interval $\Delta t$ there is a single component $E_{k}$ of a POVM $\lbrace E_{i} \rbrace$ such that $\langle U| E_{k} |U\rangle >> \langle U| E_{j} |U\rangle$ for any $j \neq k$, then the components of $|U\rangle$ for which $\lbrace E_{i} \rbrace$ yields non-negligible measures, i.e. the components of the recognized system $S^{E}$, will be superselected into the eigenstate $|S^{E}_{k}\rangle$ of $E_{k}$.  This superselection of $|S^{E}_{k}\rangle$ by the action of $\lbrace E_{i} \rbrace$ is effectively einselection: it requires during the interval $\Delta t$ both that the physical interaction between the observer and $U$ that implements the action of $\lbrace E_{i} \rbrace$ dominates the internal evolution of $S^{E}$, and that the evolution of the degrees of freedom outside $S^{E}$ have negligible effect on $|S^{E}_{k}\rangle$.  If $\lbrace E_{i} \rbrace$ is not a projective measurement, the superselected state $|S^{E}_{k}\rangle$ is only an approximate eigenstate, as is the case for einselected states if either $H_{S}$ or $H_{F}$ are non-zero.  Superselection by the action of a POVM can, therefore, be considered einselection \textit{by an observer}, as opposed to einselection by the environment.

From the perspective of an observer $O$ deploying a POVM $\lbrace E_{i} \rbrace$, the requirement for einselection that a single component $E_{k}$ exists such that $\langle U| E_{k} |U\rangle >> \langle U| E_{j} |U\rangle$ for any $j \neq k$ can be re-expressed as the requirement that for all $j \neq k$, the value $\langle U| E_{j} |U\rangle \leq \epsilon$, where $\epsilon$ is $O$'s detection threshold for information transmitted from the environment.  Physically, this formulation of the condition for einselection says that, within the time interval $\Delta t$, the Born-rule probability $\langle U| E_{k} |U\rangle$ of observing $|S^{E}_{k}\rangle$ is large while the Born-rule probabilities $\langle U| E_{j} |U\rangle$ for all $j \neq k$ are neglibible.  Under these conditions, $O$ will either record ``$|S^{E}_{k}\rangle$'' or will record nothing.  Hence from $O$'s perspective, the POVM component $E_{k}$ can be considered to be a mapping from a \textit{physical} quantum state $|U\rangle$ that is an approximate eigenstate of $E_{k}$ to an item of recordable classical \textit{information} ``$|S^{E}_{k}\rangle$.''  These items of information must be encoded physically by $O$ \cite{landauer:99}, but they are not themselves physical states of $U$: the requirement that $\langle U| E_{j} |U\rangle \leq \epsilon$ for all $j \neq k$ is non-linear and hence violates the superposition principle.  From the perspective of $O$, therefore, what is einselected by the action of $\lbrace E_{i} \rbrace$ is not a physical quantum state but an abstract, classical informational state.

Einselection by observers thus provides a mechanism for observer-dependent emergence in the epistemic sense defined by Wallace \cite{wallace:05}.  The only systems that an observer $O$ can detect, and hence the only systems that $O$ can find ``interesting'' or theorize about, are the systems recognized by the POVMs that $O$ is capable of deploying.  These POVMs constitute \textit{a priori} pragmatic information for $O$, as they are the tools - indeed, the only possible tools - that $O$ employs to obtain empirical information about the world.  An observer equipped with a POVM $\lbrace E_{i} \rbrace$ may not know, prior to making observations with $\lbrace E_{i} \rbrace$, the current state $|S^{E}\rangle$ of $S^{E}$, but such an observer knows what states of $S^{E}$ it is \textit{possible} to observe, as these are precisely the states projected by single, dominant components of $\lbrace E_{i} \rbrace$.  The recordable information $|S^{E}_{k}\rangle$ that emerges from $|U\rangle$ when $O$ deploys $\lbrace E_{i} \rbrace$ is thus constrained by $O$'s \textit{a priori} pragmatic information, but only depends physically on the prior state of $U$; hence epistemic emergence driven by POVMs is both observer-dependent and fully deterministic.

\section{Systems as virtual machines}

The sense of emergence defined here is familiar from classical computer science.  An observer faced with a physical device $D$ undergoing a sequence of state transitions can, given suitable measurements, interpret the behavior of $D$ as an execution trace of an algorithm $A$ on some specified input $a$.  For example, an observer can interpret a sequence of head positions and writings or erasures of characters on the tape of a Turing machine as the execution of a particular algorithm on a particular input.  Such an interpretation is an abstract semantic model of the behavior of $D$ as a computation, i.e. it specifies a \textit{virtual machine} realized by the physical device $D$ (e.g. \cite{tan:76}).  Run in reverse, such interpretations specify the semantics of machine-level programming languages.  The semantics of quantum computing languages \cite{gay:06, rudiger:07} have this form.

If observations are carried out with a POVM $\lbrace E_{i} \rbrace$ under the conditions for einselection described above, an observed sequence of state transitions has the form $... |D_{j}\rangle$ at $(t_{i}), |D_{k}\rangle$ at $(t_{i+1}), ...$, where $|D_{i}\rangle$ is the einselected information about $D$ recorded when $E_{i}$ is the dominant component of $\lbrace E_{i} \rbrace$ and $t_{i}$ is the time of the observation.  As noted above, the transitions between these states depend deterministically on the Hamiltonian $H_{U}$ and the POVM $\lbrace E_{i} \rbrace$.  However, they are not unitary; the unitary \textit{physical} propagator $e^{- \imath H_{U}(t)}$ is defined over the physical states $|U\rangle$, not over the einselected  \textit{informational} states $|D\rangle$.  The transitions $... |D_{j}\rangle$ at $(t_{i}), |D_{k}\rangle$ at $(t_{i+1}), ...$ are, therefore, \textit{virtual} transitions; as shown in Fig. 1, they can be considered as state transitions within a classical finite state machine (FSM) defined over the einselected $|D_{i}\rangle$.  As any observer deploying $\lbrace E_{i} \rbrace$ to gain information about $D$ knows only the probabilities $\langle U |E_{i}|U\rangle$ \textit{a priori}, the transitions between the $|D_{i}\rangle$ are transitions in a \textit{stochastic} finite state machine.

\psset{xunit=1cm,yunit=1cm}
\begin{pspicture}(0,0)(16,7.5)
\put(0.5,6){Virtual machine level:}
\put(0.8,3){Physical state level:}
\put(7.6,6.5){Virtual FSM}
\put(5.2,6){...}
\put(6,6.1){\vector(1,0){1}}
\put(7.2,6){$|D_{j}\rangle$}
\put(8.4,6.1){\vector(1,0){1}}
\put(9.6,6){$|D_{k}\rangle$}
\put(10.8,6.1){\vector(1,0){1}}
\put(12.2,6){...}

\put(7,4.5){$E_{j}$}
\put(7.6,3.5){\vector(0,1){2.2}}
\put(10,3.5){\vector(0,1){2.2}}
\put(10.2,4.5){$E_{k}$}

\put(8.3,3.5){$e^{- \imath H_{U}(t)}$}
\put(5.2,3){...}
\put(6,3.1){\vector(1,0){1}}
\put(7.1,3){$|U(t_{i})\rangle$}
\put(8.4,3.1){\vector(1,0){.8}}
\put(9.3,3){$|U(t_{i+1})\rangle$}
\put(11,3.1){\vector(1,0){.8}}
\put(12.2,3){...}

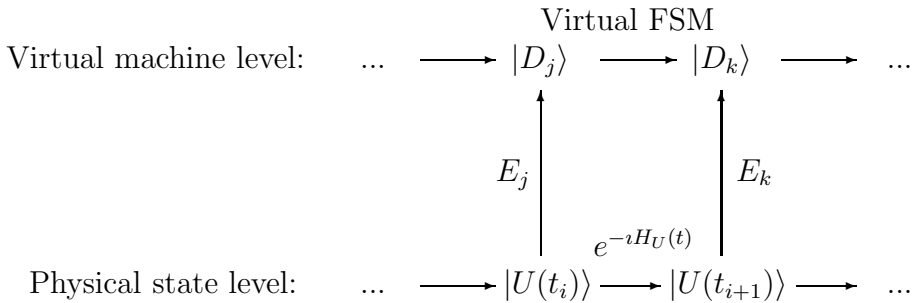
\put(0.5,1.5){\textit{Fig. 1: Semantic relationship between physical states of $U$ and einselected virtual}}
\put(0.5,1){\textit{states $|D_{i}\rangle$ specifying a virtual machine implemented by a device $D$ embedded in $U$.}}
\end{pspicture}

Treating POVM components semantically as interpretation maps as shown in Fig. 1 re-interprets ``systems'' across the board as virtual machines implemented by the elementary entities whose states are recognized by an observer-deployable POVM.  ``Systems'' are, therefore, purely informational entities that emerge in an observer-dependent, epistemic sense.  The ontological question of how systems are bounded within $\mathcal{H}_{\mathit{U}}$ evaporates; systems do not have even FAPP ontological boundaries.  Zurek's ``axiom(o)'' is no longer necessary; only the elementary entities whose degrees of freedom compose $\mathcal{H}_{\mathit{U}}$ need to be taken for granted as existing things.  All other systems exist only when someone looks, i.e. acts on $U$ with a POVM.

\section{Conclusion}

Emergence in the observer-dependent, epistemic sense defined here is a bootstrap process: from the assumption that observers can be characterized as deployers of POVMs that yield classical information, the classical information that observers obtain as a result of deploying these POVMs as information-gathering probes of the quantum world can be explained.  The explanation that is generated has, moreover, exactly the same form as is employed to explain algorithm execution by physical devices: observable physical systems are virtual machines in exactly the same sense that physically-implemented algorithms are virtual machines.  The idea that algorithm execution provides a general model for the emergent behavior of physical systems was enunciated clearly by Turing \cite{turing:50}, discussed broadly in connection with artificial intelligence (e.g. \cite{putnam:75, cummins:77, dietrich:89}), and proposed as a general account of measurement \cite{fields:89, fields:96}.  It has been overshadowed in the context of quantum theory by the goal of developing an observer-independent account of the emergence of classicality.  By showing that einselection does not provide a mechanism for observer-independent emergence, the present paper opens the way for further investigation of the conditions under which physical systems can be described as implementations of particular algorithms.

From a more practical perspective, the interpretation of ``systems'' in purely informational terms may contribute to resolving the problem of consistently describing the input and output procedures for quantum computers.  The requirement that inputs and outputs be classically interpretable and hence measurable has led to input and output procedures being treated as \textit{ad hoc} encoding and decoding steps in describing the measurement-free semantics of fully-unitary quantum computers (e.g. \cite{deutsch:85, farhi:96, nielsen-chaung:00}).  Treating classical inputs and outputs as results of measurements carried out at the two endpoints of a fully-reversible process may illuminate the somewhat problematic question of how quantum computation is to be interpreted as algorithmic \cite{aaronson:05}.

\end{document}